\documentclass[aip,jmp,12pt]{revtex4-1} 
\pdfoutput=1                 
\usepackage[utf8]{inputenc}
\usepackage{graphicx}
\newcommand{\inputpgf}[1]{\includegraphics{#1.pdf}}  

\usepackage{amsmath, amssymb}
\everymath{\displaystyle}
\renewcommand\Re{\operatorname{Re}}                                  
\DeclareMathOperator\sign{sgn}                                       
\DeclareMathOperator\Wsk{W}                                          
\DeclareMathOperator\bigo{\mathcal{O}}                               
\DeclareMathOperator\regm{\underline{M}}                             
\DeclareMathOperator\hypm{\mathit{M}}                                
\DeclareMathOperator\hypu{\mathit{U}}                                
\DeclareMathOperator\loggamma{log\Gamma}                             
\newcommand{\D}{\mathop{}\!\mathrm{d}}                               
\newcommand{\E}{\mathop{}\!\mathrm{e}}                               
\newcommand{\I}{\mathrm{i}}                                          
\newcommand{\downto}{\xrightarrow{>}}                                
\newcommand{\eps}{\varepsilon}                                       
\newcommand{\energy}{\epsilon}                                       
\newcommand{\anbohr}{a}                                              
\newcommand{\ryd}{\mathrm{Ry}}                                       
\newcommand{\der}[3][]{\frac{\D^{#1} #2}{\D #3^{#1}}}                
\newcommand{\pder}[3][]{\frac{\partial^{#1} #2}{\partial #3^{#1}}}   
\newcommand{\conj}[1]{\mkern 3mu\overline{\mkern-3mu#1}}             

\usepackage{hyperref}  
\hypersetup{
	pdfborderstyle={/S/U/W 1}, 
	pdfinfo={%
		Author       = {David Gaspard},
		Title        = {Connection formulas between Coulomb wave functions}, 
		Subject      = {Mathematical Physics},
		CreationDate = {D:20180319080000}, 
	}
}

\begin{document}
\title{Connection formulas between Coulomb wave functions} 
\author{David Gaspard}\email[E-mail:~]{dgaspard@ulb.ac.be}
\affiliation{\textsuperscript{1}Nuclear Physics and Quantum Physics, CP229, Universit\'e libre de Bruxelles (ULB), B-1050 Brussels, Belgium}
\date{\today}

\begin{abstract}
The mathematical relations between the regular Coulomb function $F_{\eta\ell}(\rho)$ and the irregular Coulomb functions $H^\pm_{\eta\ell}(\rho)$ and $G_{\eta\ell}(\rho)$ are obtained in the complex plane of the variables $\eta$ and $\rho$ for integer or half-integer values of $\ell$.
These relations, referred to as ``connection formulas'', form the basis of the theory of Coulomb wave functions, and play an important role in many fields of physics, especially in the quantum theory of charged particle scattering.
As a first step, the symmetry properties of the regular function $F_{\eta\ell}(\rho)$ are studied, in particular under the transformation $\ell\mapsto-\ell-1$, by means of the modified Coulomb function $\Phi_{\eta\ell}(\rho)$, which is entire in the dimensionless energy $\eta^{-2}$ and the angular momentum $\ell$.
Then, it is shown that, for integer or half-integer $\ell$, the irregular functions $H^\pm_{\eta\ell}(\rho)$ and $G_{\eta\ell}(\rho)$ can be expressed in terms of the derivatives of $\Phi_{\eta,\ell}(\rho)$ and $\Phi_{\eta,-\ell-1}(\rho)$ with respect to $\ell$.
As a consequence, the connection formulas directly lead to the description of the singular structures of $H^\pm_{\eta\ell}(\rho)$ and $G_{\eta\ell}(\rho)$ at complex energies in their whole Riemann surface.
The analysis of the functions is supplemented by novel graphical representations in the complex plane of $\eta^{-1}$.
\end{abstract}
\maketitle

\section{Introduction\label{sec:Introduction}}
The Coulomb wave functions are defined as particular solutions of the Schrödinger equation in a $1/r$ potential.
They have been introduced in the 1930s by Yost, Wheeler and Breit~\cite{Yost1936} to describe the scattering of charged particles due to the Coulomb repulsion.
Most of the properties of these functions have been developed by Breit and Hull,\cite{Breit1950a, Breit1950b, Breit1959} Abramowitz and Stegun,\cite{Abramowitz1954, Stegun1955, Abramowitz1964} and later by Seaton,\cite{Seaton1982, Seaton1983, Seaton2002a} Thompson and Barnett,\cite{Thompson1985, Thompson1986} Humblet~\cite{Humblet1964, Humblet1984, Humblet1985, Humblet1990} and others.
As pointed out in Ref.~\citenum{Dzieciol1999}, the literature about the Coulomb functions is so broad and plentiful that it is often difficult to find formulas covering general properties, such as the relations between the regular and the irregular Coulomb functions, their analytic continuation to complex arguments, or their singular structures in the complex plane of the energy.
These important properties are rarely given in reference handbooks,\cite{Olver2010, DLMF} yet equivalent formulas are given for the closely related Bessel functions or the confluent hypergeometric functions.
\par However, the Coulomb wave functions for complex arguments are of major interest in various fields such as charged particle scattering in nuclear and atomic physics,\cite{Humblet1984, Humblet1990, Lambert1969, Hamilton1973, Joachain1979, Newton1982, Haeringen1985, Rakityansky2013} quantum defect theory,\cite{Seaton1983} Regge pole theory,\cite{Takemasa1979} or the scattering of gravitational waves.\cite{Asada1997, Berti2006c}
The in-depth survey of these functions has led to recent advances in the theory of effective-range function for charged particles.\cite{GaspardD2018a, RamirezSuarez2017, Blokhintsev2017, Blokhintsev2018}
\par The main purpose of this paper is to determine the relations between the regular and the irregular Coulomb functions, respectively denoted as $F_{\eta\ell}(\rho)$ and $G_{\eta\ell}(\rho)$, in an easy-to-read fashion. 
These relations are generically referred to as the ``connection formulas'' in the NIST Handbook.\cite{Olver2010}
The focus of this paper is on integer or half-integer values of the angular momentum $\ell$ and complex values of the variables $\eta$ and $\rho$.
When $\ell$ is an integer or a half-integer, the connection formulas relate $G_{\eta\ell}(\rho)$ to derivatives of $F_{\eta\ell}(\rho)$ with respect to the angular momentum $\ell$.\cite{Humblet1984, Dzieciol1999}  
The connection formulas are of prime importance since they allow the user to deduce from any representation of $F_{\eta\ell}(\rho)$ valid for complex $\ell$, the corresponding representation of $G_{\eta\ell}(\rho)$.
The properties of the resulting representation of $G_{\eta\ell}(\rho)$ are also inherited from $F_{\eta\ell}(\rho)$.\cite{Humblet1984}
\par This paper investigates the limitation of $F_{\eta\ell}(\rho)$ and $G_{\eta\ell}(\rho)$ at complex energies due to their singularities, and introduces regularized versions of the Coulomb functions, namely $\Phi_{\eta\ell}(\rho)$ and $\Psi_{\eta\ell}(\rho)$, that are holomorphic in energy.
It is shown why these versions are more suitable than $F_{\eta\ell}(\rho)$ and $G_{\eta\ell}(\rho)$ for the analysis and the complex continuation of Coulomb-related functions, especially at low energy.
\par On this ground, it is shown that the connection formulas directly lead to the analysis of the singularities of $G_{\eta\ell}(\rho)$ in the complex plane of the energy, as well as the construction of its Riemann surface.
The analysis is also illustrated by novel color-based representations of the functions in the complex plane of $\eta^{-1}$.\cite{Wegert2012}
\par This paper is organized as follows.
The main definitions and notations are given in Sec.~\ref{sec:Coulomb-reminder}.
The important symmetry properties of $F_{\eta\ell}(\rho)$ are shown in Sec.~\ref{sec:Coulomb-symmetries}, especially the transformation $\ell\mapsto-\ell-1$ which is discussed in Sec.~\ref{sec:Coulomb-symmetry-index}.
The connection formulas between the regular and the irregular Coulomb functions are presented in Sec.~\ref{sec:Coulomb-connections}, in addition to the analytic continuation, symmetries, and other little known properties of $G_{\eta\ell}(\rho)$.
The results for complex angular momentum are given in Sec.~\ref{sec:Coulomb-connections-1}.
Finally, the special case of integer and half-integer values of $\ell$ is treated in Sec.~\ref{sec:Coulomb-connections-2}.
Conclusions are drawn in Sec.~\ref{sec:Conclusions}.

\section{Coulomb equation and standard solutions\label{sec:Coulomb-reminder}}
\par The Coulomb wave functions arise while solving the Schrödinger equation for a spinless particle of mass $m$ in a stationary $1/r$ potential.
Using the radial coordinate $\rho=kr$ rescaled by the wave number $k$, the Schrödinger equation of the wave function $u(\rho)$ reads
\begin{equation}\label{eq:Coulomb-eq}
-\der[2]{u}{\rho} + \left[\frac{\ell(\ell+1)}{\rho^2} + \frac{2\eta}{\rho} - 1\right] u = 0  \:.
\end{equation}
The Sommerfeld parameter $\eta$ quantifies the distortion of the wave function due to the Coulomb potential with respect to the free wave of corresponding angular momentum $\ell$. For instance, when $\eta$ tends to zero, the radial Schrödinger equation in free space is retrieved.
The parameter $\eta$ is commonly defined as~\cite{Yost1936, Seaton1982, Seaton2002a, Olver2010, Joachain1979, Newton1982, Haeringen1985}
\begin{equation}\label{eq:Sommerfeld-param}
\eta = \frac{1}{\anbohr k} = \frac{1}{\sqrt{\energy}}  \:,
\end{equation}
where $\energy$ is the dimensionless energy in Rydberg units
\begin{equation}\label{eq:Dimensionless-energy}
\energy = \eta^{-2} = (\anbohr k)^2 = \frac{E}{\ryd}  \:,
\end{equation}
and $\anbohr$ denotes the Bohr radius.
\par In this paper, the focus is on the analytic properties of the functions in the complex planes of $k$ and $\energy$.
In this regard, it will be convenient to let the product $\eta\rho$ appear in place of the radial coordinate, because, being equal to $r/\anbohr$, it does not depend on $k$ or $\energy$ anymore.
\par In the literature,\cite{Yost1936, Abramowitz1964, Olver2010, Joachain1979, Newton1982, Haeringen1985} two couples of linearly independent functions are considered as solutions of the Coulomb equation~\eqref{eq:Coulomb-eq}, namely $\{F_{\eta\ell},G_{\eta\ell}\}$ and $\{H^+_{\eta\ell},H^-_{\eta\ell}\}$.
The former consists of the regular Coulomb function $F_{\eta\ell}(\rho)$, behaving like $\rho^{\ell+1}$ as $\rho\rightarrow 0$, and the irregular Coulomb function $G_{\eta\ell}(\rho)$, behaving like $\rho^{-\ell}$ as $\rho\downto 0$.
It can be shown that the only solution up to a factor of the Coulomb equation~\eqref{eq:Coulomb-eq} which is regular at $\rho=0$ is given by~\cite{Abramowitz1964, Olver2010}
\begin{equation}\label{eq:Coulomb-F}
F_{\eta\ell}(\rho) = C_{\eta\ell}\,\rho^{\ell+1}\E^{\pm\I\rho} \hypm(\ell+1\pm\I\eta, 2\ell+2, \mp2\I\rho)  \:,
\end{equation}
where $\hypm(\alpha,\beta,z)$ is the confluent hypergeometric function $_1F_1$, also known as the Kummer function of the first kind.
The Kummer function is defined by the following series representation with $\alpha=\ell+1\pm\I\eta$, $\beta=2\ell+2$, and $z=\mp2\I\rho$,
\begin{equation}\label{eq:Kummer-M}
\hypm(\alpha, \beta, z) = \sum_{n=0}^\infty \frac{(\alpha)_n}{(\beta)_n}\frac{z^n}{n!} = 1 + \frac{\alpha}{\beta}\frac{z}{1!} + \frac{\alpha(\alpha+1)}{\beta(\beta+1)}\frac{z}{2!} + \ldots  \:,
\end{equation}
where $(\alpha)_n = \alpha(\alpha+1)\cdots(\alpha+n-1) = \Gamma(\alpha+n)/\Gamma(\alpha)$ is the Pochhammer symbol.
The choice of signs in the definition~\eqref{eq:Coulomb-F} is immaterial due to Kummer's reflection formula of the confluent hypergeometric function.\cite{Abramowitz1964, Olver2010}
It should be noted that the definition~\eqref{eq:Coulomb-F} is supposed to hold for complex $\ell$.
The properties of the Coulomb functions owing to the symmetry formulas of the confluent hypergeometric functions are discussed in further details in Sec.~\ref{sec:Coulomb-symmetries}.
\par In Eq.~\eqref{eq:Coulomb-F}, the coefficient $C_{\eta\ell}$ normalizes the far-field behavior of $F_{\eta\ell}$ to a sine wave of unit amplitude
\begin{equation}\label{eq:Far-field-F}
F_{\eta\ell}(\rho) \longrightarrow \sin\theta_{\eta\ell}(\rho) \quad\mathrm{as}~\rho\rightarrow\infty  \:,
\end{equation}
where $\theta_{\eta\ell}$ is the far-field phase of the Coulomb wave functions~\cite{Abramowitz1964, Olver2010}
\begin{equation}\label{eq:Coulomb-phase}
\theta_{\eta\ell}(\rho) = \rho - \ell\frac{\pi}{2} - \eta\ln(2\rho) + \arg\Gamma(\ell+1+\I\eta)  \:.
\end{equation}
The normalization coefficient is defined as~\cite{Humblet1984, Humblet1985, Abramowitz1964, Olver2010}
\begin{equation}\label{eq:Coulomb-C}
C_{\eta\ell} = \frac{2^\ell\sqrt{\Gamma(\ell+1+\I\eta)\Gamma(\ell+1-\I\eta)}}{\Gamma(2\ell+2)\E^{\eta\pi/2}}  \:.
\end{equation}
Other definitions of $C_{\eta\ell}$ valid in the complex plane of the wave number $k$ are discussed in Sec.~\ref{sec:Coulomb-symmetries}.
\par
\begin{figure*}[ht]
\centering\inputpgf{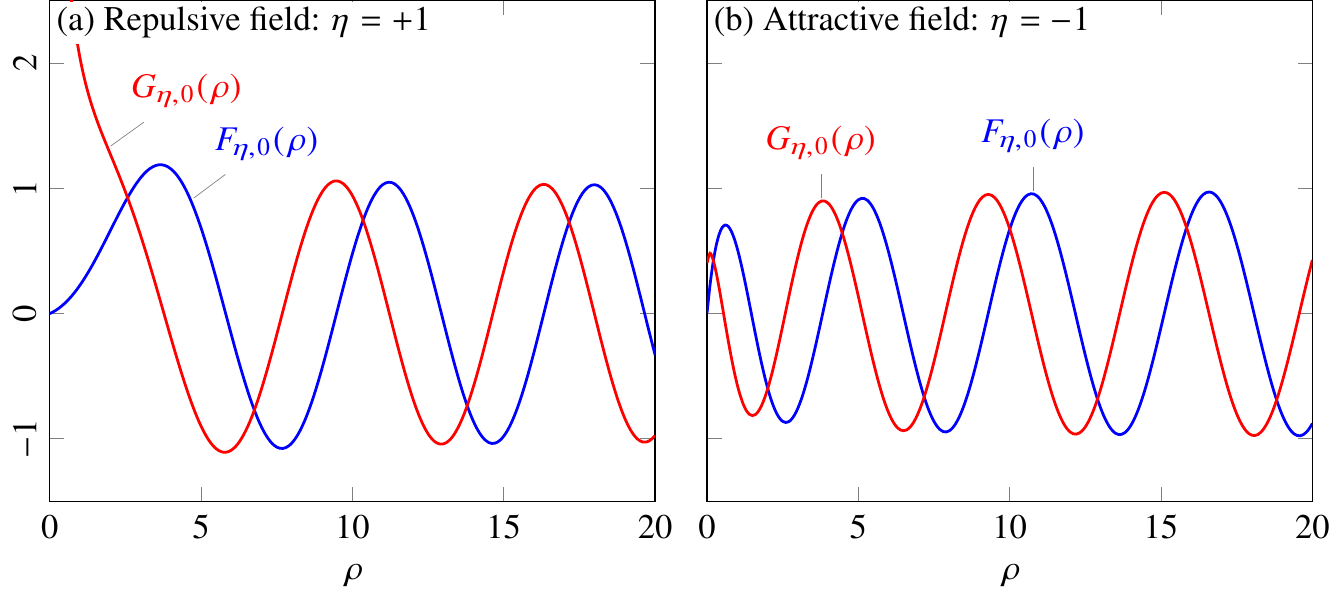}
\caption{\label{fig:plot-coulomb-functions}Graphs of the standard Coulomb wave functions $F_{\eta\ell}(\rho)$ and $G_{\eta\ell}(\rho)$ for $\eta=\pm1$ and $\ell=0$. At the origin, $G_{\eta,0}(\rho)$ is equal to the finite value $G_{\eta,0}(0)=1/C_{\eta,0}$.}
\end{figure*}

 The regular Coulomb function $F_{\eta\ell}(\rho)$ is shown in Fig.~\ref{fig:plot-coulomb-functions} in a repulsive and an attractive field.
This is the only function which vanishes at $\rho=0$.%
\par The definition of the irregular Coulomb function $G_{\eta\ell}$ is less straightforward than $F_{\eta\ell}$.
As a preamble, additional Coulomb functions have to be introduced, namely the outgoing and incoming Coulomb functions denoted $H^+_{\eta\ell}$ and $H^-_{\eta\ell}$ respectively.
These two functions are defined in a similar way to Eq.~\eqref{eq:Coulomb-F},
\begin{equation}\label{eq:Coulomb-H}
H^\pm_{\eta\ell}(\rho) = D^\pm_{\eta\ell}\,\rho^{\ell+1}\E^{\pm\I\rho} \hypu(\ell+1\pm\I\eta, 2\ell+2, \mp2\I\rho)  \:,
\end{equation}
but using the confluent hypergeometric function of the second kind $\hypu(\alpha,\beta,z)$, also known as the Tricomi function~\cite{Abramowitz1964, Olver2010}
\begin{equation}\label{eq:Tricomi-U-1}
\hypu(\alpha,\beta,z) = \frac{\Gamma(1-\beta)}{\Gamma(\alpha-\beta+1)}\hypm(\alpha, \beta, z) + \frac{\Gamma(\beta-1)}{\Gamma(\alpha)}z^{1-\beta}\hypm(\alpha-\beta+1, 2-\beta, z)
\end{equation}
After inspection of Eqs.~\eqref{eq:Coulomb-H} and~\eqref{eq:Tricomi-U-1}, one notices that $H^\pm_{\eta\ell}$ can be directly expressed as a combination of $F_{\eta,\ell}$ and $F_{\eta,-\ell-1}$, and can thus be defined in this way.\cite{Buchholz1969, Dzieciol1999}
These considerations are discussed in Sec.~\ref{sec:Coulomb-connections}.
\par The functions $H^\pm_{\eta\ell}$ are irregular near $\rho=0$, and behave like $\rho^{-\ell}$ as $\rho\downto 0$.\cite{Abramowitz1964, Olver2010}
The normalization coefficients $D^\pm_{\eta\ell}$ in Eq.~\eqref{eq:Coulomb-H} are defined as~\cite{Olver2010}
\begin{equation}\label{eq:Coulomb-D}
D^\pm_{\eta\ell} = (\mp2\I)^{2\ell+1} \frac{\Gamma(\ell+1\pm\I\eta)}{C_{\eta\ell}\Gamma(2\ell+2)}  \:,
\end{equation}
such that the far-field behavior of $H^\pm_{\eta\ell}$ is $\E^{\pm\I\theta_{\eta\ell}}$.
The coefficients $D^\pm_{\eta\ell}$ are related to each other by complex conjugation as long as $k$ is real.
The general conjugation formulas for complex valued $k$ read
\begin{equation}\label{eq:Coulomb-conjugate}
D^-_{\eta\ell} = \conj{D^+_{\conj{\eta}\ell}} \quad\mathrm{and}\quad H^-_{\eta\ell}(kr) = \conj{H^+_{\conj{\eta}\ell}(\conj{k}r)}  \:.
\end{equation}
The regular Coulomb function $F_{\eta\ell}$ can be retrieved by taking the imaginary part of $H^+_{\eta\ell}$ in the following way
\begin{equation}\label{eq:Coulomb-F-2}
F_{\eta\ell}(\rho) = \frac{H^+_{\eta\ell}(\rho) - H^-_{\eta\ell}(\rho)}{2\I}  \:.
\end{equation}
This definition is consistent with the expected far-field behavior~\eqref{eq:Far-field-F} of $F_{\eta\ell}$.
Finally, the irregular Coulomb function $G_{\eta\ell}$ is defined as the real part of $H^+_{\eta\ell}$
\begin{equation}\label{eq:Coulomb-G}
G_{\eta\ell}(\rho) = \frac{H^+_{\eta\ell}(\rho) + H^-_{\eta\ell}(\rho)}{2}  \:,
\end{equation}
so that $G_{\eta\ell}$ asymptotically behaves like $\cos\theta_{\eta\ell}$ as $\rho\rightarrow\infty$.\cite{Abramowitz1964, Olver2010, Michel2007}
\par The function $G_{\eta\ell}$ is shown along with $F_{\eta\ell}$ in Fig.~\ref{fig:plot-coulomb-functions} for $\ell=0$.
As one can see, $G_{\eta,0}$ is irregular at $\rho=0$, but it has no vertical asymptote, contrary to what Fig.~\ref{fig:plot-coulomb-functions}(a) might suggest.
In fact, $G_{\eta,0}$ tends to a positive constant for $\rho\downto0$ according to the following behavior~\cite{Abramowitz1964, Olver2010}
\begin{equation}\label{eq:Near-field-G}
G_{\eta\ell}(\rho) = \frac{\rho^{-\ell}}{(2\ell+1)C_{\eta\ell}} + \left\{\begin{array}{ll}%
\bigo(\rho\ln\rho)       & \mathrm{if}~\ell=0     \\
\bigo(\rho^{-\ell+1})    & \mathrm{if}~\ell>0 \end{array}\right.  \:.
\end{equation}
When $\ell=0$, the next-to-leading order in the series expansion of $G_{\eta\ell}$ is exceptionally logarithmic, hence the conditions in \eqref{eq:Near-field-G}.

\section{Symmetries of Coulomb functions\label{sec:Coulomb-symmetries}}
\par Before deriving the connection formulas between Coulomb wave functions, we have to discuss their symmetries with respect to their parameters.
The connection formulas between functions are based on the symmetries of the differential equation they satisfy.
For instance, the Schrödinger equation~\eqref{eq:Coulomb-eq} shows a noticeable symmetry in the angular momentum: it is left unchanged under the transformation $\ell\mapsto-\ell-1$.
Therefore, one can guess that connection formulas between Coulomb functions for $\ell$ and for $-\ell-1$ should exist, in particular between the regular and the irregular Coulomb functions.
\par To obtain these formulas, we first have to define modified Coulomb functions that are regular for any $\ell\in\mathbb{C}$ in the complex plane of the energy.
This step is motivated by the singularities of the Kummer function $\hypm(\alpha,\beta,z)$ in the complex $\beta$ plane at $\beta\in\mathbb{Z}_{\leq0}$.
As a reminder, $\beta$ equals $2\ell+2$ and thus can reach negative integer values when $\ell$ is changed to $-\ell-1$.
Therefore, we need the regularized version $\regm(\alpha,\beta,z)$ of the Kummer function which is simultaneously holomorphic in the complex planes of $\alpha$, $\beta$ and $z$:~\cite{Olver2010}
\begin{equation}\label{eq:Kummer-regularized}
\regm(\alpha,\beta,z) = \frac{1}{\Gamma(\beta)}\hypm(\alpha,\beta,z) = \sum_{n=0}^\infty \frac{(\alpha)_n}{\Gamma(\beta+n)}\frac{z^n}{n!}  \:.
\end{equation}
On the basis of $\regm(\alpha,\beta,z)$, it is useful to introduce a modified Coulomb function $\Phi_{\eta\ell}$ satisfying Eq.~\eqref{eq:Coulomb-eq}
\begin{equation}\label{eq:Coulomb-Phi}
\boxed{\Phi_{\eta\ell}(\rho) = \left(2\eta\rho\right)^{\ell+1} \E^{\pm\I\rho} \regm(\ell+1\pm\I\eta, 2\ell+2, \mp2\I\rho)}  \:, 
\end{equation}
which is holomorphic in the complex planes of the wave number $k$, the energy $\energy$, and $\ell$.
Similar functions are also reported in the literature.\cite{Yost1936, Humblet1984, Humblet1985, Rakityansky2013}
In Refs.~\citenum{Seaton1982, Seaton1983, Seaton2002a}, Seaton denotes $\Phi_{\eta\ell}$ as $(-1)^{\ell+1}f(\energy,\ell;-r/\anbohr)$.
\par Such a function is of major interest in charged particle scattering,\cite{Joachain1979, Newton1982, Haeringen1985} Regge pole theory~\cite{Takemasa1979} and quantum defect theory,\cite{Seaton1983} for which analytic functions in the complex plane of the energy are often required.\cite{Seaton2002a, Dzieciol1999}
In particular, the function $\Phi_{\eta\ell}$ behaves like a constant in the neighborhood of the zero-energy point~\cite{Abramowitz1964, Olver2010}:
\begin{equation}\label{eq:Coulomb-Phi-low-energy}
\Phi_{\eta\ell}(\rho) = x\,I_{2\ell+1}(2x) + \bigo(\energy)   \quad\mathrm{as}~\energy\rightarrow0  \:,
\end{equation}
where $x=\sqrt{2\eta\rho}=\sqrt{2r/\anbohr}$ and $I_{\nu}(x)$ is the modified Bessel function of the first kind.
It should be noted that the low-energy behavior~\eqref{eq:Coulomb-Phi-low-energy} is valid for both repulsive ($\anbohr>0$) and attractive ($\anbohr<0$) Coulomb potentials.
In the latter case, the variable $x$ becomes imaginary due to the square root.
However, the result is still real valued because $x\,I_{2\ell+1}(2x)$ is an even function of $x$ provided $\ell\in\mathbb{Z}$.
\par The modified Coulomb function $\Phi_{\eta\ell}$ is related to $F_{\eta\ell}$ by
\begin{equation}\label{eq:Coulomb-F-from-Phi}
F_{\eta\ell}(\rho) = \frac{C_{\eta\ell}\Gamma(2\ell+2)}{(2\eta)^{\ell+1}} \Phi_{\eta\ell}(\rho)  \:.
\end{equation}
The normalization factor in Eq.~\eqref{eq:Coulomb-F-from-Phi} brings many singularities to $F_{\eta\ell}$ in the complex planes of $\eta^{-1}$ and $\ell$, which makes $F_{\eta\ell}$ less suited than $\Phi_{\eta\ell}$ for our purposes.

\subsection{Transformations of the main variables\label{sec:Coulomb-symmetry-variables}}
\par When the Coulomb interaction is attractive, $\Phi_{\eta\ell}$ reduces to the hydrogen-like wave function within a normalization factor.
Under the transformation $(\anbohr,k)\mapsto(-\anbohr,\I k)$, or in other words $(\eta,\rho)\mapsto(\I\eta,\I\rho)$, the hypergeometric function in Eq.~\eqref{eq:Coulomb-Phi} can be rewritten as a generalized Laguerre polynomial $L^{(\alpha)}_n(x)$.
Then, we have
\begin{equation}
\Phi_{\I\eta,\ell}(\I\rho) = \frac{\Gamma(\eta-\ell)}{\Gamma(\eta+\ell+1)} (-2\eta\rho)^{\ell+1}\E^{-\rho} L^{(2\ell+1)}_{\eta-\ell-1}(2\rho)  \:,
\end{equation}
where $\eta$ is interpreted as the principal quantum number belonging to positive integers $\eta\in\mathbb{Z}_{>0}$, $\ell\in\{0, 1, \ldots, \eta-1\}$, and $\rho=r/(\anbohr\eta)$.
\par As previously mentioned, the immaterial choice of signs in Eqs.~\eqref{eq:Coulomb-F} and~\eqref{eq:Coulomb-Phi} is a consequence of the Kummer reflection formula of the confluent hypergeometric function~\cite{Abramowitz1964, Olver2010}
\begin{equation}\label{eq:Kummer-reflection-1}
\regm(\alpha,\beta,z) = \E^{z}\regm(\beta-\alpha,\beta,-z)  \quad\forall\alpha,\beta,z\in\mathbb{C}  \:,
\end{equation}
which remains valid in the complex planes of $\alpha$, $\beta$ and $z$ as well.
\par Another important consequence of Eq.~\eqref{eq:Kummer-reflection-1} is the relation between the repulsive and the attractive Coulomb functions.
One easily shows from Eq.~\eqref{eq:Kummer-reflection-1} that the attractive Coulomb function is found on the negative real $r$ axis:
\begin{equation}\label{eq:Coulomb-Phi-attractive}
\Phi_{-\eta,\ell}(\rho) = \Phi_{\eta,\ell}(-\rho)  \quad\forall\rho,\eta,\ell\in\mathbb{C}  \:.
\end{equation}
This property can also be interpreted as the consequence of the invariance of the Schrödinger equation~\eqref{eq:Coulomb-eq} under the transformation $(\eta,\rho)\mapsto(-\eta,-\rho)$.
A similar relation also exists for the Coulomb function $F_{\eta\ell}$:~\cite{Dzieciol1999, Michel2007}
\begin{equation}\label{eq:Coulomb-F-attractive}
F_{-\eta,\ell}(\rho) = -\E^{\pi(\eta\pm\I\ell)} F_{\eta,\ell}(\rho\E^{\mp\I\pi})  \quad\mathrm{for}~\pm\arg\rho>0  \:.
\end{equation}
While Eq.~\eqref{eq:Coulomb-Phi-attractive} is valid on the complex planes of the three variables, the relation~\eqref{eq:Coulomb-F-attractive} suffers from restrictions because of the normalization coefficient in the definition~\eqref{eq:Coulomb-F-from-Phi}.
When $\ell$ is not an integer, the Coulomb functions $F_{\eta\ell}(\rho)$ and $F_{\eta\ell}(-\rho)$ display different branch cuts in the complex $\rho$ plane: the former lies on the negative $\rho$ axis and the latter on the positive $\rho$ axis.
Therefore, the principal branches of $F_{\eta\ell}(\rho)$ and $F_{\eta\ell}(-\rho)$ cannot be proportional within the same factor everywhere in the complex $\rho$ plane.
The upper sign in Eq.~\eqref{eq:Coulomb-F-attractive} holds for $0<\arg\rho\leq\pi$ and the lower sign for $-\pi<\arg\rho\leq0$.  

\subsection{Reflection of the angular momentum\label{sec:Coulomb-symmetry-index}}
\par The reflection formula~\eqref{eq:Kummer-reflection-1} is not the only symmetry of the confluent hypergeometric function $\regm(\alpha,\beta,z)$.
There is another formula involving the integral parameter $\beta$,\cite{Bateman1953, Lambert1969} namely
\begin{equation}\label{eq:Kummer-reflection-2}
\regm(\alpha-\beta+1,2-\beta,z) = \frac{z^{\beta-1}\,\Gamma(\alpha)}{\Gamma(\alpha-\beta+1)}\regm(\alpha,\beta,z) \quad\forall\beta\in\mathbb{Z}  \:.
\end{equation}
This property derives from the series representation~\eqref{eq:Kummer-regularized} under the assumption that $\beta$ is an integer.
One notices that the transformations $\alpha\mapsto\alpha-\beta+1$ and $\beta\mapsto2-\beta$ involved in Eq.~\eqref{eq:Kummer-reflection-2} are equivalent to the symmetry $\ell\mapsto-\ell-1$ leaving the Schrödinger equation~\eqref{eq:Coulomb-eq} unchanged.
Therefore, it is useful to determine the equivalent relation in terms of the Coulomb functions.
For this purpose, one can define two functions with the prefactor in the right-hand side of Eq.~\eqref{eq:Kummer-reflection-2}
\begin{equation}\label{eq:Coulomb-w-plus}
w^\pm_{\eta\ell} = \frac{\Gamma(\ell+1\pm\I\eta)}{(\pm\I\eta)^{2\ell+1}\,\Gamma(-\ell\pm\I\eta)}  \:.
\end{equation}
These functions are denoted as $A(\energy,\ell)$ by Seaton.\cite{Seaton1982, Seaton1983, Seaton2002a, Olver2010}
For positive integer $\ell$, the two functions $w^\pm_{\eta\ell}$ are entire in the energy $\energy$ and both reduce to $w_{\eta\ell}$, the $\ell$-order polynomial in $\energy$ given by~\cite{Breit1950b, Breit1959, Humblet1984, Humblet1985, GaspardD2018a, RamirezSuarez2017, Baye2000a}  
\begin{equation}\label{eq:Coulomb-w}
w^\pm_{\eta\ell} = w_{\eta\ell} = \prod_{j=0}^\ell \left(1 + \frac{j^2}{\eta^2}\right)  \quad\forall\ell\in\mathbb{Z}_{\geq0}  \:.
\end{equation}
These functions $w^\pm_{\eta\ell}$ and $w_{\eta\ell}$ are equal to $1$ if $\energy=0$ or $\ell=0$.
The formula~\eqref{eq:Coulomb-w} can be supplemented by the reflection property
\begin{equation}\label{eq:Coulomb-w-reflection}
w^\pm_{\eta,-\ell-1} = \frac{1}{w^\pm_{\eta,\ell}}  \quad\forall\ell\in\mathbb{C}  \:.
\end{equation}
\par It should be noted that the reflection formula~\eqref{eq:Kummer-reflection-2} is also valid for half-odd-integer $\ell$ since $\beta=2\ell+2~\in\mathbb{Z}$.
In that case, the functions $w^\pm_{\eta\ell}$ and $w_{\eta\ell}$ read
\begin{equation}
w^\pm_{\eta\ell} = w_{\eta\ell} = \prod_{j=1/2}^\ell \left(1 + \frac{j^2}{\eta^2}\right)  \quad\forall\ell\in\{\tfrac{1}{2}, \tfrac{3}{2}, \tfrac{5}{2}, \ldots\}  \:,
\end{equation}
where the index $j$ runs through half odd integers until $\ell$ is reached: $j\in\{\tfrac{1}{2}, \tfrac{3}{2}, \tfrac{5}{2}, \ldots, \ell\}$.
Most of the relations derived in this paper allow $\ell$ to be half a odd integer (possibly negative), although these values have no physical meaning in the framework of Coulomb scattering.
For this reason, the extension to half-integer values of $\ell$ is rarely considered in the literature.\cite{Dzieciol1999}
\par The reflection formula of $\Phi_{\eta\ell}$ is obtained by multiplying both sides of Eq.~\eqref{eq:Kummer-reflection-2} by $(2\eta\rho)^{\ell+1}\E^{\pm\I\rho}$.
One gets
\begin{equation}\label{eq:Coulomb-Phi-reflection}
\boxed{\Phi_{\eta,-\ell-1}(\rho) = w_{\eta\ell}\,\Phi_{\eta,\ell}(\rho) \quad\forall\ell\in\tfrac{1}{2}\mathbb{Z}}  \:,
\end{equation}
where $\tfrac{1}{2}\mathbb{Z}$ stands for the set of all half-integers: $\{0,\pm\tfrac{1}{2},\pm1,\pm\tfrac{3}{2},\ldots\}$.
This result plays a key role in the derivation of the connection formulas for the irregular Coulomb functions in Sec.~\ref{sec:Coulomb-connections}.
\par
\begin{figure*}[ht]
\centering\inputpgf{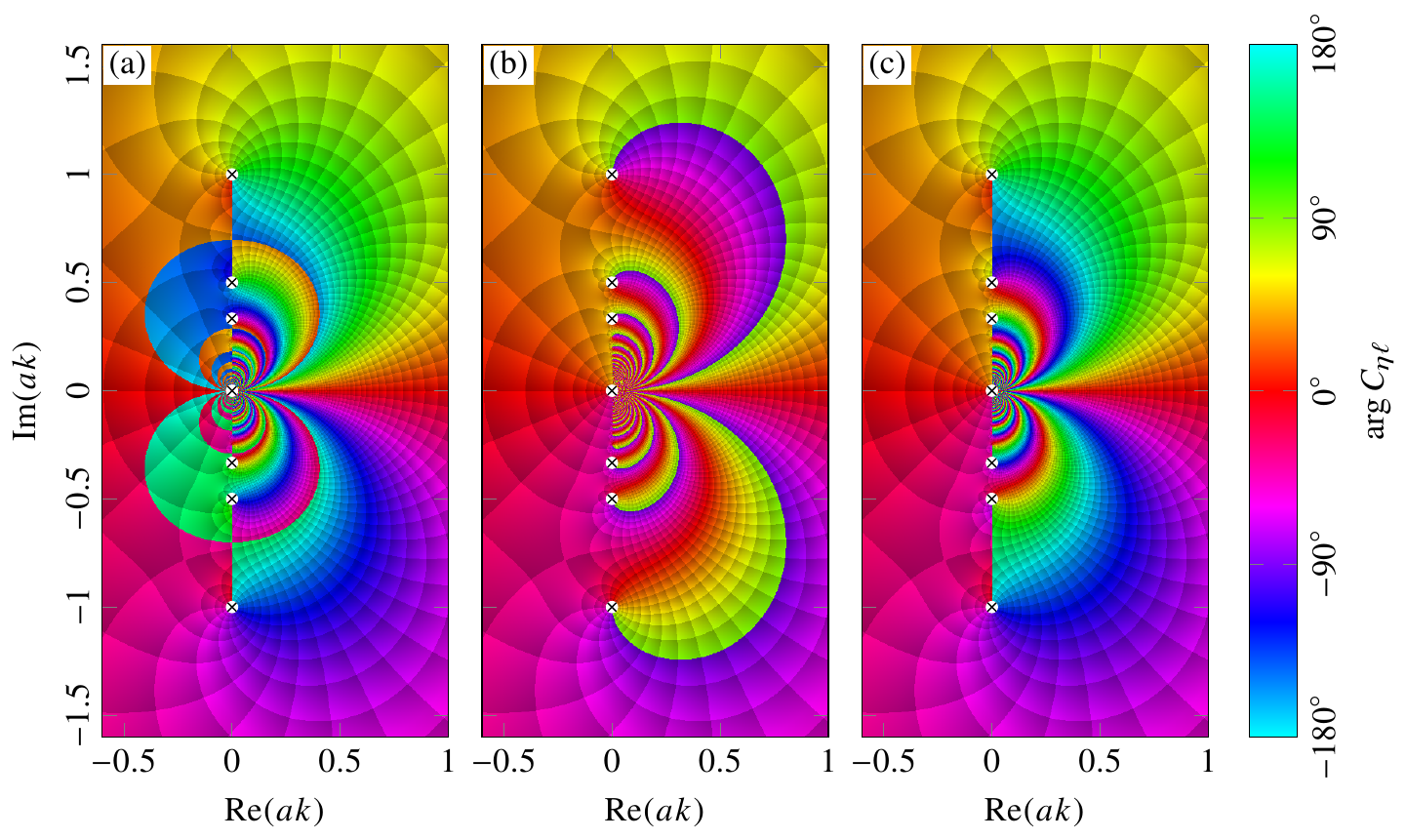}
\caption{\label{fig:map-Coulomb-C}(Color online) Phase plots~\cite{Wegert2012} of the Coulomb normalization coefficient $C_{\eta\ell}$ in the complex plane of $\eta^{-1}=\anbohr k$ for $\ell=0$ using three different definitions: (a) from Eq.~\eqref{eq:Coulomb-C} with the square root of gamma functions, (b) from Eq.~\eqref{eq:Coulomb-C-from-w} and (c) from Eq.~\eqref{eq:Coulomb-C-LogGamma} using definition~\eqref{eq:LogGamma} of $\loggamma$. Some of the branch points located at $\anbohr k = \pm\I/(n+\ell+1)~\forall n\in\mathbb{Z}_{\geq0}$ are marked with crosses.
The point $\anbohr k=0$ is an essential singularity.}
\end{figure*}

 At this point, it is crucial to understand the relationship between the symmetries of $\Phi_{\eta\ell}$ and $F_{\eta\ell}$.
With this aim, the reflection formula~\eqref{eq:Coulomb-Phi-reflection} should be rewritten in terms of the standard Coulomb function $F_{\eta\ell}$ using Eq.~\eqref{eq:Coulomb-F-from-Phi}.
However, this step is less easy than deriving Eq.~\eqref{eq:Coulomb-Phi-reflection} for $\Phi_{\eta\ell}$ because the normalization coefficient $C_{\eta\ell}$, which will come into play, is a multi-valued function of the wave number $k$ and has no conventional principal branch in the complex plane of $k$.
In fact, the coefficient $C_{\eta\ell}$ defined by Eq.~\eqref{eq:Coulomb-C} has a tangled structure of branch cuts, as shown in Fig.~\ref{fig:map-Coulomb-C}(a), due to the square root on the gamma functions.
The branch points of $C_{\eta\ell}$ are given by its poles located at $\anbohr k = \pm\I/(n+\ell+1)~\forall n\in\mathbb{Z}_{\geq0}$.
These points are referred to as the ``Coulomb poles'' because they are reminiscent of the hydrogen-like spectrum, $n$ being interpreted as the radial quantum number.%
\par Depending on the definition of $C_{\eta\ell}$, different branch cut structures can be obtained.
For instance, using Euler's reflection formula of the gamma function, we get the alternate form of $C_{\eta\ell}$
\begin{equation}\label{eq:Coulomb-C-from-w}
C_{\eta\ell} = \frac{(2\eta)^\ell}{\Gamma(2\ell+2)}\sqrt{\frac{2\eta\pi\,w_{\eta\ell}^\pm}{\E^{2\eta\pi} - \E^{\mp2\pi\I\ell}}}  \quad\forall\ell\in\mathbb{C}  \:,
\end{equation}
where the choice of the upper or the lower sign is immaterial as far as the branch cuts of $w^+_{\eta\ell}$ and $w^-_{\eta\ell}$ coincide.
The corresponding branch cut structure is shown in Fig.~\ref{fig:map-Coulomb-C}(b).
Because of the square root in Eq.~\eqref{eq:Coulomb-C-from-w}, all the Coulomb poles are joined to the origin $\anbohr k=0$ by a branch cut in the right half-plane.
It can be guessed from Fig.~\ref{fig:map-Coulomb-C} that these different forms for $C_{\eta\ell}$ define the same multifunction.
Indeed, we see that the branch cuts have been displaced from Fig.~\ref{fig:map-Coulomb-C}(a) to~\ref{fig:map-Coulomb-C}(b).
\par In order to understand the symmetry properties of the coefficient $C_{\eta\ell}\Gamma(2\ell+2)/(2\eta)^{\ell+1}$ relating the functions $F_{\eta\ell}$ and $\Phi_{\eta\ell}$ in Eq.~\eqref{eq:Coulomb-F-from-Phi}, one has to formulate $C_{\eta\ell}$ so as to minimize the length of the branch cuts.
For this purpose, one defines the log-gamma function for $N\rightarrow\infty$ as~\cite{Kolbig1972, Michel2007}
\begin{equation}\label{eq:LogGamma}
\loggamma(z) = \left[(z+N)(\ln(z+N)-1) - \frac{1}{2}\ln\frac{z+N}{2\pi}\right] - \sum_{n=0}^{N-1}\ln(z+n)  \:.
\end{equation}
The log-gamma function is equal to $\ln(\Gamma(z))$ for $z\in\mathbb{R}_{>0}$, but it has a single discontinuity for $z\in\mathbb{R}_{<0}$ stemming from the superposition of the logarithmic branch cuts in the sum.
In this regard, the log-gamma function has a simpler branch cut structure than $\ln(\Gamma(z))$.
The expression~\eqref{eq:LogGamma} originates from the recurrence formula $\loggamma(z+1)=\loggamma(z)+\ln z$ and the asymptotic Stirling expansion of the gamma function~\cite{Abramowitz1964, Olver2010} displayed between square brackets.
The latter can be supplemented by additional terms in the Stirling expansion to speed up the convergence of the expression.
\par Now, using the log-gamma function, $C_{\eta\ell}$ can be rewritten from Eq.~\eqref{eq:Coulomb-C} as~\cite{Michel2007}
\begin{equation}\label{eq:Coulomb-C-LogGamma}
C_{\eta\ell} = \frac{2^\ell\E^{-\eta\pi/2}}{\Gamma(2\ell+2)} \exp\left(\frac{\loggamma(\ell+1+\I\eta) + \loggamma(\ell+1-\I\eta)}{2}\right)  \:.
\end{equation}
The coefficient $C_{\eta\ell}$ defined by Eqs.~\eqref{eq:Coulomb-C-LogGamma} and~\eqref{eq:LogGamma} is shown in Fig.~\ref{fig:map-Coulomb-C}(c).
The branch cut structure turns out to be an alternation of branch cuts between the Coulomb poles.
Indeed, it turns out to be shorter and less tangled than in Figs.~\ref{fig:map-Coulomb-C}(a) or~\ref{fig:map-Coulomb-C}(b).
Unless otherwise stated, we will use Eq.~\eqref{eq:Coulomb-C-LogGamma} to compute $C_{\eta\ell}$ for complex valued wave numbers $k$.
\par
\begin{figure*}[ht]
\centering\inputpgf{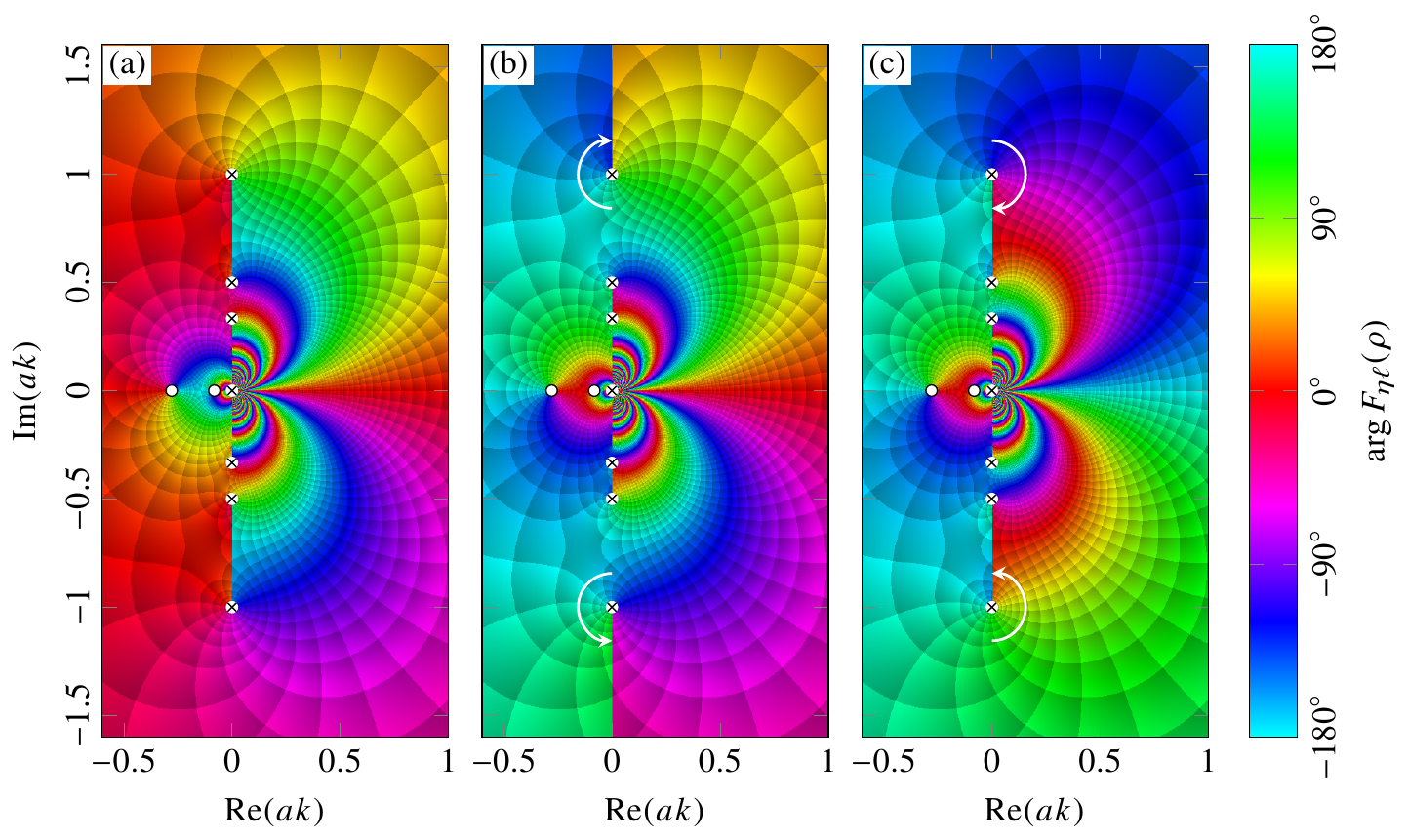}
\caption{\label{fig:map-Coulomb-F}(Color online) Phase plots~\cite{Wegert2012} of $F_{\eta\ell}(\rho)$ in the complex plane of $\anbohr k$ at $\rho=1/2$ for $\ell=0$ using definition~\eqref{eq:Coulomb-C-LogGamma} of $C_{\eta\ell}$.
Panel (a) depicts the principal branch of $F_{\eta,\ell}$, panel (b) shows $F_{\eta,-\ell-1}$, and panel (c) is the second branch of $F_{\eta,\ell}$, namely $-F_{\eta,\ell}$.
The first nontrivial zeroes are depicted by circles.}
\end{figure*}

 Returning to the reflection formula of $F_{\eta\ell}$, we multiply both sides of Eq.~\eqref{eq:Coulomb-Phi-reflection} by $C_{\eta\ell}\Gamma(2\ell+2) (2\eta)^{-\ell-1}(w_{\eta\ell})^{-1}$ and use Eq.~\eqref{eq:Coulomb-F-from-Phi} to get
\begin{equation}\label{eq:Coulomb-F-reflection-1}  
\frac{C_{\eta\ell}\Gamma(2\ell+2)}{(2\eta)^{\ell+1}w_{\eta\ell}}\Phi_{\eta,-\ell-1}(\rho) = F_{\eta,\ell}(\rho) \quad\forall\ell\in\tfrac{1}{2}\mathbb{Z}  \:.
\end{equation}
Now, for complex values of $\ell$, some manipulations involving the definitions~\eqref{eq:Coulomb-C-LogGamma} of $C_{\eta\ell}$ and~\eqref{eq:Coulomb-w-plus} of $w^\pm_{\eta\ell}$ show that the coefficient in the left hand side of Eq.~\eqref{eq:Coulomb-F-reflection-1} reduces to the coefficient in front of $\Phi_{\eta\ell}$ in Eq.~\eqref{eq:Coulomb-F-from-Phi} with $\ell$ replaced by $-\ell-1$.
Accordingly, one finds the following symmetry for the coefficient in front of $\Phi_{\eta,-\ell-1}$ in Eq.~\eqref{eq:Coulomb-F-reflection-1}:
\begin{equation}\label{eq:Coulomb-C-reflection-1}
\frac{C_{\eta,\ell}\Gamma(2\ell+2)}{(2\eta)^{\ell+1}w^\pm_{\eta\ell}} = \frac{C_{\eta,-\ell-1}\Gamma(-2\ell)}{(2\eta)^{-\ell}} \sqrt{\frac{w^\mp_{\eta\ell}}{w^\pm_{\eta\ell}}}  \quad\forall\ell\in\mathbb{C} \:. 
\end{equation}
The gamma functions in $\sqrt{w^\mp_{\eta\ell}/w^\pm_{\eta\ell}}$ are supposed to be computed in the exponential form using Eq.~\eqref{eq:LogGamma}.
It can be shown\cite{Dzieciol1999} that the last factor in Eq.~\eqref{eq:Coulomb-C-reflection-1} is equivalent to the sign of the real part of $\eta$ whatever the sign over the functions $w^\pm_{\eta\ell}$:
\begin{equation}\label{eq:Coulomb-C-reflection-2}
\sqrt{\frac{w^\mp_{\eta\ell}}{w^\pm_{\eta\ell}}} = (\pm\I)^{2\ell+1}\sqrt{\frac{\Gamma(\ell+1\mp\I\eta)\Gamma(-\ell\pm\I\eta)}{\Gamma(\ell+1\pm\I\eta)\Gamma(-\ell\mp\I\eta)}} = \sign(\Re\eta)^{2\ell+1}  \quad\forall\ell\in\tfrac{1}{2}\mathbb{Z}  \:.
\end{equation}
Therefore, combining Eqs.~\eqref{eq:Coulomb-F-reflection-1}, \eqref{eq:Coulomb-C-reflection-1}, and~\eqref{eq:Coulomb-C-reflection-2}, we obtain the following reflection formula of the regular Coulomb function also reported in Ref.~\citenum{Dzieciol1999} 
\begin{equation}\label{eq:Coulomb-F-reflection-2}
F_{\eta,-\ell-1}(\rho) = \sign(\Re\eta)^{2\ell+1}\,F_{\eta,\ell}(\rho)  \quad\forall\ell\in\tfrac{1}{2}\mathbb{Z}  \:.
\end{equation}
As shown in Figs.~\ref{fig:map-Coulomb-F}(a) and~\ref{fig:map-Coulomb-F}(b), the functions $F_{\eta,\ell}$ and $F_{\eta,-\ell-1}$ are equal in the right half-plane ($\Re\eta>0$).
However, their extension to the left half-plane ($\Re\eta<0$) is different because the branch cuts have been rotated by $180^\circ$ around the Coulomb poles, as depicted by the arrows.  
In Fig.~\ref{fig:map-Coulomb-F}(b), one notices that the branch cuts split the complex plane in two along the imaginary axis. 
The resulting discontinuity is responsible for the sign of $\Re\eta$ in Eq.~\eqref{eq:Coulomb-F-reflection-2}.%
\par Considering the Riemann structure of $F_{\eta\ell}$, it is clear from Figs.~\ref{fig:map-Coulomb-F}(a), (b), and (c) that $F_{\eta\ell}$, $F_{\eta,-\ell-1}$, and $-F_{\eta\ell}$ are different branches of the same multifunction $\mathbf{F}_{\eta\ell}$ defined as:
\begin{equation}\label{eq:Coulomb-F-surface}
\mathbf{F}_{\eta\ell}(\rho) = \{\pm F_{\eta\ell}(\rho)\}  \quad\forall\ell\in\tfrac{1}{2}\mathbb{Z} \:.
\end{equation}
This two-sheeted structure is due to the square root in the Coulomb normalization coefficient $C_{\eta\ell}$ of Eq.~\eqref{eq:Coulomb-C-from-w}.
In terms of the multifunction $\mathbf{F}_{\eta\ell}$, the reflection formula~\eqref{eq:Coulomb-F-reflection-2} becomes:
\begin{equation}\label{eq:Coulomb-F-reflection-3}
\mathbf{F}_{\eta,-\ell-1}(\rho) = \mathbf{F}_{\eta,\ell}(\rho)  \quad\forall\ell\in\tfrac{1}{2}\mathbb{Z}  \:.
\end{equation}
\par The conclusion is that the use of the Coulomb function $F_{\eta\ell}$ is limited due to its singular structure at complex energy.
Indeed, the symmetry relation~\eqref{eq:Coulomb-F-reflection-2} involves the discontinuous coefficient $\sign(\Re\eta)$, in contrast to the symmetry~\eqref{eq:Coulomb-Phi-reflection} of $\Phi_{\eta\ell}$ whose coefficient $w_{\eta\ell}$ is entire in energy.
This is why the modified function $\Phi_{\eta\ell}$ is more suitable than $F_{\eta\ell}$ for practical applications such as the analysis of Coulomb-related functions.
In particular, the connection formula relating $F_{\eta\ell}$ and $G_{\eta\ell}$ and the analysis of $G_{\eta\ell}$ will be more easily discussed in terms of $\Phi_{\eta\ell}$ rather than $F_{\eta\ell}$ itself.

\section{Connection formulas of Coulomb functions\label{sec:Coulomb-connections}}
In this section, we establish the connection formulas between the Coulomb wave functions $F_{\eta\ell}$, $G_{\eta\ell}$ and $H^\pm_{\eta\ell}$ valid in repulsive and attractive electric fields.
In this regard, we use the modified Coulomb function $\Phi_{\eta\ell}$ which are regular in the complex plane of the energy.
We first discuss the general case where $\ell$ is complex in Sec.~\ref{sec:Coulomb-connections-1}.
Then, we study the specialization to the integer and half-integer values of $\ell$ in Sec.~\ref{sec:Coulomb-connections-2}.
We show that the use of $\Phi_{\eta\ell}$ directly leads to the analytic decomposition of the irregular Coulomb function $G_{\eta\ell}$ which is a key component in the effective-range theory for charged particle scattering.\cite{Humblet1990, Rakityansky2013, GaspardD2018a}

\subsection{Complex angular momentum\label{sec:Coulomb-connections-1}}
As mentioned in Sec.~\ref{sec:Coulomb-reminder}, the irregular Coulomb functions are based on the Tricomi function $\hypu(\alpha,\beta,z)$.
Replacing the confluent hypergeometric functions $\hypm(\alpha,\beta,z)$ in Eq.~\eqref{eq:Tricomi-U-1} with its regularized version $\regm(\alpha,\beta,z)$ defined by Eq.~\eqref{eq:Kummer-regularized}, one gets
\begin{equation}\label{eq:Tricomi-U-2}
\hypu(\alpha,\beta,z) = \Gamma(\beta)\Gamma(1-\beta)\!\left[\frac{\regm(\alpha,\beta,z)}{\Gamma(\alpha-\beta+1)} - \frac{\regm(\alpha-\beta+1,2-\beta,z)}{z^{\beta-1}\,\Gamma(\alpha)}\right]  \!,
\end{equation}
which has a form directly related to the symmetry formula~\eqref{eq:Kummer-reflection-2}.
When $\beta=2\ell+2$ is not an integer, a suitable function linearly independent from $\regm(\alpha,\beta,z)$ can be $z^{1-\beta}\regm(\alpha-\beta+1,2-\beta,z)$.\cite{Abramowitz1964, Olver2010, Lambert1969, Bateman1953}
However, when $\beta$ tends to an integer, the property~\eqref{eq:Kummer-reflection-2} breaks this linear independence.
In the latter case, the Tricomi function~\eqref{eq:Tricomi-U-2} is needed because it remains linearly independent for $\ell\in\tfrac{1}{2}\mathbb{Z}$.
The poles of the prefactor $\Gamma(\beta)\Gamma(1-\beta)$ compensate for the cancellation of the square brackets when $\beta$ reaches an integer.
Multiplying Eq.~\eqref{eq:Tricomi-U-2} by $D^\pm_{\eta\ell}\,\rho^{\ell+1}\E^{\pm\I\rho}$ to let appear the Coulomb functions $H^\pm_{\eta\ell}$ and $\Phi_{\eta\ell}$ and using the definition~\eqref{eq:Coulomb-D} of $D^\pm_{\eta\ell}$, one finds
\begin{equation}\label{eq:Coulomb-H-from-Phi-0}
\boxed{H^\pm_{\eta\ell} = \frac{(2\eta)^\ell}{C_{\eta\ell}\Gamma(2\ell+2)} \,\frac{\pi}{\sin(2\pi\ell)} \left[w^\pm_{\eta,\ell}\Phi_{\eta,\ell} - \Phi_{\eta,-\ell-1}\right]  \quad\forall\ell\in\mathbb{C}}  \:.
\end{equation}
In fact, the expression~\eqref{eq:Coulomb-H-from-Phi-0} provides the analytic decomposition of $H^\pm_{\eta\ell}$ in the complex planes of $\eta^{-1}$ and $\ell$, since the singularities are gathered in $C_{\eta\ell}$ and $w^\pm_{\eta\ell}$.
In particular, $C_{\eta\ell}$ and $(2\eta)^\ell$ are multivalued functions for complex $\ell$.
A similar decomposition can be obtained for $G_{\eta\ell}$ using Eq.~\eqref{eq:Coulomb-G}.
\par The expression~\eqref{eq:Coulomb-H-from-Phi-0} can also be reformulated in terms of the Coulomb function $F_{\eta\ell}$ as
\begin{equation}\label{eq:Coulomb-H-from-F-0}
H^\pm_{\eta\ell} = \frac{(2\eta)^{2\ell+1}}{C_{\eta\ell}^2\Gamma(2\ell+2)^2} \,\frac{\pi}{\sin(2\pi\ell)} \left[w^\pm_{\eta\ell}F_{\eta,\ell} - \left(w^+_{\eta\ell}w^-_{\eta\ell}\right)^{1/2}F_{\eta,-\ell-1}\right]  \:.
\end{equation}
It turns out that the coefficient in front of the square brackets in Eq.~\eqref{eq:Coulomb-H-from-F-0} can be simplified further.
Using Eq.~\eqref{eq:Coulomb-C-from-w}, we notice that the following imaginary part of $w^+_{\eta\ell}$ reduces to the denominator in Eq.~\eqref{eq:Coulomb-H-from-F-0}
\begin{equation}
\frac{w^+_{\eta\ell}-w^-_{\eta\ell}}{2\I} = \frac{C_{\eta\ell}^2\Gamma(2\ell+2)^2}{(2\eta)^{2\ell+1}}\frac{\sin(2\pi\ell)}{\pi}  \quad\forall\ell\in\mathbb{C}  \:.
\end{equation}
Therefore, the Coulomb functions $H^\pm_{\eta\ell}$ can be rewritten as
\begin{equation}\label{eq:Coulomb-H-from-F-gen}
H^\pm_{\eta\ell} = \frac{w^\pm_{\eta\ell}F_{\eta,\ell} - \left(w^+_{\eta\ell}w^-_{\eta\ell}\right)^{1/2}F_{\eta,-\ell-1}}{\tfrac{1}{2\I}\left(w^+_{\eta\ell}-w^-_{\eta\ell}\right)}  \:.
\end{equation}
According to the definition~\eqref{eq:Coulomb-G} of $G_{\eta\ell}$, the real part of Eq.~\eqref{eq:Coulomb-H-from-F-gen} gives
\begin{equation}\label{eq:Coulomb-G-from-F-gen}
G_{\eta\ell} = \frac{\tfrac{1}{2}\left(w^+_{\eta\ell}+w^-_{\eta\ell}\right)F_{\eta,\ell} - \left(w^+_{\eta\ell}w^-_{\eta\ell}\right)^{1/2}F_{\eta,-\ell-1}}{\tfrac{1}{2\I}\left(w^+_{\eta\ell}-w^-_{\eta\ell}\right)}  \:.
\end{equation}
The result~\eqref{eq:Coulomb-G-from-F-gen} can be easily related to Eq.~(6.1) of Ref.~\citenum{Dzieciol1999} using the notation $\E^{2\I x_{\eta\ell}}=w^-_{\eta\ell}/w^+_{\eta\ell}$.
The real and imaginary parts of $(w^-_{\eta\ell}/w^+_{\eta\ell})^{1/2}$ are then expressed as $\cos x_{\eta\ell}$ and $\sin x_{\eta\ell}$ respectively.\cite{Dzieciol1999}
This notation highlights the analogy between Eq.~\eqref{eq:Coulomb-G-from-F-gen} and the connection formula between the Bessel functions $J_\nu$ and $Y_\nu$ for $\nu\in\mathbb{C}$, as given in Refs.~\citenum{Rade2004, Olver2010}.
\par Finally, as a consequence of Eq.~\eqref{eq:Coulomb-H-from-Phi-0}, one finds using Eqs.~\eqref{eq:Coulomb-w-reflection} and~\eqref{eq:Coulomb-C-reflection-1} that the Coulomb function $H^\pm_{\eta\ell}$ obeys the symmetry formula\cite{Dzieciol1999, Buchholz1969, Bateman1953}
\begin{equation}\label{eq:Coulomb-H-reflection-gen}
H^\pm_{\eta,-\ell-1} = \sqrt{\frac{w^\mp_{\eta\ell}}{w^\pm_{\eta\ell}}} H^\pm_{\eta\ell}  \quad\forall\ell\in\mathbb{C}  \:,
\end{equation}
where the square root of gamma functions can be computed using the log-gamma function from Eq.~\eqref{eq:LogGamma}.
In contrast to the symmetry formula~\eqref{eq:Coulomb-Phi-reflection} for $\Phi_{\eta\ell}$ which is restricted to $\ell\in\tfrac{1}{2}\mathbb{Z}$, the result~\eqref{eq:Coulomb-H-reflection-gen} is valid for all complex values of $\ell$.
\par It should be noted that the result~\eqref{eq:Coulomb-G-from-F-gen} can be extracted from Eq.~\eqref{eq:Coulomb-H-reflection-gen} by subtracting the equations with the upper and lower signs and solving for $G_{\eta\ell}$.
However, in general, there is no symmetry formula strictly equivalent to Eq.~\eqref{eq:Coulomb-H-reflection-gen} for the functions $F_{\eta\ell}$ and $G_{\eta\ell}$.
The reason is that the multiplicative function $\sqrt{w^\mp_{\eta\ell}/w^\pm_{\eta\ell}}$ is generally a complex number that couples the components of $F_{\eta\ell}$ and $G_{\eta\ell}$ in $H^\pm_{\eta\ell}$, preventing Eq.~\eqref{eq:Coulomb-H-reflection-gen} from splitting into two independent symmetry formulas for $F_{\eta\ell}$ and $G_{\eta\ell}$.
An important exception occurs when $\ell$ is either integer or half-integer, as shown in the following.

\subsection{Integer or half-integer angular momentum\label{sec:Coulomb-connections-2}}
\par Now, we consider the special case of integer or half-integer values of $\ell$.
This case is not only important for physical applications, it also requires a careful mathematical treatment of the limit on $\ell$.
The calculation of the limit is easier on Eq.~\eqref{eq:Coulomb-H-from-Phi-0} than on Eq.~\eqref{eq:Coulomb-H-from-F-gen} because of the known behavior of the sine function at each of its zeroes.
In addition, since $\Phi_{\eta\ell}$ is regular in $\ell$, the calculation will provide the analytic decomposition of $H^\pm_{\eta\ell}$ in the special case $\ell\in\tfrac{1}{2}\mathbb{Z}$. 
Letting $\ell$ be replaced by $\ell+\eps$ for $\ell\in\tfrac{1}{2}\mathbb{Z}$ with vanishing $\eps$, the connection formula~\eqref{eq:Coulomb-H-from-Phi-0} becomes
\begin{equation}\label{eq:Coulomb-H-from-Phi-1}
H^\pm_{\eta\ell} = \frac{(2\eta)^\ell(-1)^{2\ell}}{C_{\eta\ell}\Gamma(2\ell+2)} \;\lim_{\eps\rightarrow0}\frac{1}{2\eps}\left[w^\pm_{\eta,\ell+\eps}\Phi_{\eta,\ell+\eps} - \Phi_{\eta,-\ell-\eps-1}\right]  \quad\forall\ell\in\tfrac{1}{2}\mathbb{Z}\:.
\end{equation}
In this limit, the square brackets in Eq.~\eqref{eq:Coulomb-H-from-Phi-1} can be related to derivatives with respect to $\ell$ by l'Hospital's rule, as done in Ref.~\citenum{Buchholz1969}.
In the following, these derivatives will be denoted by the dot notation for convenience\cite{Humblet1984}
\begin{equation}\label{eq:Coulomb-DotPhi}
\dot{\Phi}_{\eta,\ell} = \pder{\Phi_{\eta,\ell}}{\ell}  \:,
\end{equation}
and similarly for $\dot{w}^\pm_{\eta\ell}$ and other functions of $\ell$.
Therefore, the subtraction in Eq.~\eqref{eq:Coulomb-H-from-Phi-1} is expanded at first order in $\eps$ as
\begin{equation}\label{eq:Coulomb-H-bracket-1}\arraycolsep=0pt\begin{array}{ll}%
 & w^\pm_{\eta,\ell+\eps}\Phi_{\eta,\ell+\eps} - \Phi_{\eta,-\ell-\eps-1}  \\
 & = (w_{\eta\ell} + \eps\dot{w}^\pm_{\eta\ell}) (\Phi_{\eta\ell} + \eps\dot{\Phi}_{\eta\ell}) - (\Phi_{\eta,-\ell-1} - \eps\dot{\Phi}_{\eta,-\ell-1}) + \bigo(\eps^2) \:.
\end{array}\end{equation}
Remarkably, the functions $\dot{w}^\pm_{\eta\ell}$ in Eq.~\eqref{eq:Coulomb-H-bracket-1} are closely related to another well-known function in the framework of charged particle scattering,\cite{Humblet1984, Humblet1985, Humblet1990, Lambert1969, Hamilton1973, GaspardD2018a, RamirezSuarez2017} namely the $h$ function defined as~\cite{Seaton1982, Seaton2002a}
\begin{equation}\label{eq:Bethe-h}
h^\pm_{\eta\ell} = \frac{\dot{w}^\pm_{\eta\ell}}{2w_{\eta\ell}} = \frac{\psi(\ell+1\pm\I\eta) + \psi(-\ell\pm\I\eta)}{2} - \ln(\pm\I\eta)  \:,  
\end{equation}
where $\psi(z)$ is the digamma function defined as $\psi(z) = \Gamma'(z)/\Gamma(z)$.\cite{Abramowitz1964, Olver2010}
The result~\eqref{eq:Bethe-h} directly follows from the definition~\eqref{eq:Coulomb-w-plus}.
\par The functions $h^\pm_{\eta\ell}$ are more commonly defined as independent from $\ell$, i.e., removing $\ell$ in Eq.~\eqref{eq:Bethe-h}.
The dependence in $\ell$ is generally separated from $h^\pm_{\eta\ell}$ by means of the recurrence formula of the digamma function.
Although that separation is widely encountered in the literature,\cite{Humblet1984, Humblet1985, Humblet1990, Hamilton1973, Rakityansky2013} it does not lead to any subsequent simplifications in the resulting connection formulas.
This is why we prefer the definition~\eqref{eq:Bethe-h} as in Refs.~\citenum{Seaton1982, Seaton2002a, GaspardD2018a}.%
\begin{figure*}[ht]
\centering\inputpgf{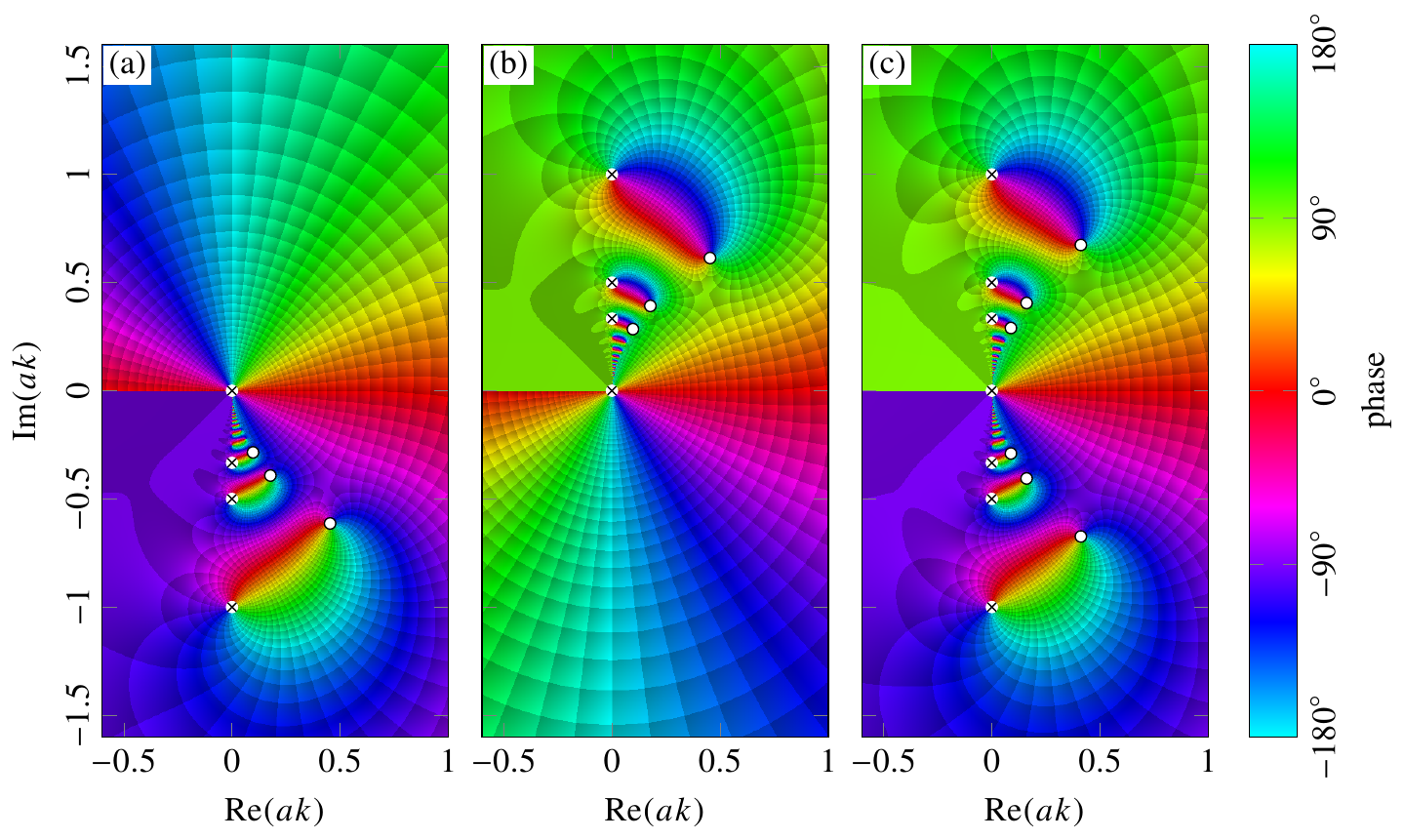}
\caption{\label{fig:map-Bethe-g}(Color online) Phase plots~\cite{Wegert2012} of the functions $h^\pm_{\eta\ell}$ and $g_{\eta\ell}$ in the complex plane of $\anbohr k$ for $\ell=0$.
The functions $h^+_{\eta\ell}$ and $h^-_{\eta\ell}$ are shown in (a) and (b) respectively, and $g_{\eta\ell}$ in (c).
Every branch cut has been turned towards the negative real axis by continuation of the logarithm.
The functions are slowly varying at $\Re(\anbohr k)<0$ near the Coulomb poles, hence the uniform color.}
\end{figure*}%
\par At this point, it is important to note that the functions $h^\pm_{\eta\ell}$ are singular in the complex plane of the wave number $k$ contrary to the Coulomb function $\Phi_{\eta\ell}$.
They exhibit the Coulomb poles accumulating to the origin in addition to a logarithmic branch cut from $\anbohr k=0$ to infinity, as shown in Figs.~\ref{fig:map-Bethe-g}(a) and~\ref{fig:map-Bethe-g}(b).
These functions $h^\pm_{\eta\ell}$ play an important role in the theory of the effective-range function in charged particle scattering.\cite{Breit1959, Humblet1964, Humblet1990, Hamilton1973, Rakityansky2013}
\par Using the reflection formula~\eqref{eq:Coulomb-Phi-reflection} and the functions $h^\pm_{\eta\ell}$ from Eq.~\eqref{eq:Bethe-h}, the result~\eqref{eq:Coulomb-H-bracket-1} reduces at first order in $\eps$ to
\begin{equation}\label{eq:Coulomb-H-bracket-2}
w^\pm_{\eta,\ell+\eps}\Phi_{\eta,\ell+\eps} - \Phi_{\eta,-\ell-\eps-1}
 = \eps\left[w_{\eta\ell}\dot{\Phi}_{\eta\ell} + \dot{\Phi}_{\eta,-\ell-1} + 2w_{\eta\ell}h^\pm_{\eta\ell}\Phi_{\eta\ell}\right] + \bigo(\eps^2)  \:.
\end{equation}
\par Additional simplifications come from the remaining normalization factor in Eq.~\eqref{eq:Coulomb-H-from-Phi-1}.
From Eq.~\eqref{eq:Coulomb-C-from-w}, it can be expressed as
\begin{equation}\label{eq:Coulomb-H-coefficient}
\frac{(2\eta)^\ell(-1)^{2\ell}}{C_{\eta\ell}\Gamma(2\ell+2)} = \frac{\E^{2\pi(\eta+\I\ell)}-1}{\pi\,w_{\eta\ell}}\,\frac{C_{\eta\ell}\Gamma(2\ell+2)}{(2\eta)^{\ell+1}}  \:.  
\end{equation}
The idea behind this expression is the appearance of the coefficient between $F_{\eta\ell}$ and $\Phi_{\eta\ell}$ in Eq.~\eqref{eq:Coulomb-F-from-Phi}.
Therefore, from Eqs.~\eqref{eq:Coulomb-H-from-Phi-1}, \eqref{eq:Coulomb-H-bracket-2}, and~\eqref{eq:Coulomb-H-coefficient}, the irregular Coulomb functions $H^\pm_{\eta\ell}$ can be rewritten
\begin{equation}\label{eq:Coulomb-H-from-Phi-2}
H^\pm_{\eta\ell} = \frac{\E^{2\pi(\eta+\I\ell)}-1}{\pi\,w_{\eta\ell}}\,\frac{C_{\eta\ell}\Gamma(2\ell+2)}{(2\eta)^{\ell+1}}\left[\frac{w_{\eta\ell}}{2}\dot{\Phi}_{\eta,\ell} + \frac{1}{2}\dot{\Phi}_{\eta,-\ell-1} + w_{\eta\ell}h^\pm_{\eta\ell}\Phi_{\eta\ell}\right]  \:.
\end{equation}
Since the derivatives $\dot{\Phi}_{\eta,\ell}$ and $\dot{\Phi}_{\eta,-\ell-1}$ are the only functions linearly independent from $\Phi_{\eta\ell}$ in Eq.~\eqref{eq:Coulomb-H-from-Phi-2}, it is useful to define another irregular solution of the Schrödinger equation~\eqref{eq:Coulomb-eq}
\begin{equation}\label{eq:Coulomb-Psi}
\boxed{\Psi_{\eta\ell}(\rho) = \frac{w_{\eta\ell}}{2}\dot{\Phi}_{\eta,\ell}(\rho) + \frac{1}{2}\dot{\Phi}_{\eta,-\ell-1}(\rho)  \quad\forall\ell\in\tfrac{1}{2}\mathbb{Z}}  \:.
\end{equation}
The function $\Psi_{\eta\ell}$ has the noticeable property of being holomorphic in the complex plane of the wave number $k$ or the energy $\energy$, especially in the neighborhood of $\energy=0$.\cite{Humblet1984, Lambert1969}
It is worth noting that $\Psi_{\eta\ell}$ is solely defined for integer or half-integer $\ell$, and has no generalization to complex-valued~$\ell$.
Indeed, the function $\Psi_{\eta\ell}$ is not expected to satisfy the Schr\"{o}dinger equation~\eqref{eq:Coulomb-eq} for $\ell\in\mathbb{C}\setminus\tfrac{1}{2}\mathbb{Z}$.
\par The relation~\eqref{eq:Coulomb-Psi} can be looked upon as the generalization to the Coulomb functions of the connection formulas between Bessel functions.
In fact, using Eqs.~\eqref{eq:Coulomb-Phi-low-energy}, \eqref{eq:Coulomb-Psi}, and 10.27.5 of Ref.~\citenum{Olver2010}, one immediately gets the limiting behavior:
\begin{equation}\label{eq:Coulomb-Psi-low-energy}
\Psi_{\eta\ell}(\rho) = 2x\,K_{2\ell+1}(2x) + \bigo(\energy)  \quad\mathrm{as}~\energy\rightarrow 0  \:,
\end{equation}
where $x=\sqrt{2r/\anbohr}$ and $K_\nu(x)$ is the modified Bessel function of the second kind.
It should be noted that the irregular Coulomb functions $H^\pm_{\eta\ell}$ and $G_{\eta\ell}$ also tend to be proportional to $2x\,K_{2\ell+1}(2x)$ as $\energy\downto 0$,\cite{Olver2010} because the functions $h^\pm_{\eta\ell}$ in Eq.~\eqref{eq:Coulomb-H-from-Phi-2} are asymptotic to zero as $\energy\downto 0$.
However, this limit only holds for $\energy>0$, since $h^\pm_{\eta\ell}$ display an essential singularity at $\energy=0$, as shown in Fig.~\ref{fig:map-Bethe-g}.
\par The appearance of derivatives with respect to $\ell$ in Eq.~\eqref{eq:Coulomb-Psi} is specific to the Coulomb potential.
When the Coulomb interaction vanishes ($\eta=0$), $F_{\eta\ell}$ and $G_{\eta\ell}$ reduce to spherical Bessel functions.
In this special case, these functions are directly related to each other by the transformation $\ell\mapsto-\ell-1$ without derivative with respect to $\ell$:\cite{Olver2010}
\begin{equation}
G_{0,\ell}(\rho) = (-1)^{\ell}F_{0,-\ell-1}(\rho)  \quad\forall\ell\in\mathbb{Z}  \:. 
\end{equation}%
\par On the basis of $\Psi_{\eta\ell}$, one defines a last irregular Coulomb function $I_{\eta\ell}$ similarly to Eq.~\eqref{eq:Coulomb-F-from-Phi}:
\begin{equation}\label{eq:Coulomb-I-from-Psi}
I_{\eta\ell}(\rho) = \frac{C_{\eta\ell}\Gamma(2\ell+2)}{(2\eta)^{\ell+1}}\Psi_{\eta\ell}(\rho)  \:.
\end{equation}
This function is real for $\rho>0$ and satisfies the Schrödinger equation~\eqref{eq:Coulomb-eq}.
The function $I_{\eta\ell}$ is the same as in Ref.~\citenum{GaspardD2018a}, and has the advantage of being on an equal footing with $F_{\eta\ell}$ regarding the $k$ dependence.
Indeed, both $F_{\eta\ell}$ and $I_{\eta\ell}$ behave as $C_{\eta\ell}k^{\ell+1}$ in the neighborhood of $k=0$.
Therefore, they are not holomorphic in $k$ or $\energy$.
In this regard, the newly introduced functions $\Phi_{\eta\ell}$ and $\Psi_{\eta\ell}$ form a couple of independent solutions that are holomorphic in $k$ and $\energy$.
The Wronskian with respect to the radial coordinate $2r/\anbohr$ reads
\begin{equation}
\Wsk_{2r/\anbohr}\!\left[\Psi_{\eta\ell}, \Phi_{\eta\ell}\right] = 1  \quad\mathrm{for}~\anbohr^{-1}\neq 0  \:,
\end{equation}
where $\Wsk_x[f,g] = f\partial_xg - g\partial_xf$.
This result also confirms that $\Psi_{\eta\ell}$ is holomorphic in $k$ and $\energy$.
\par At last, using Eq.~\eqref{eq:Coulomb-H-from-Phi-2} and the relations~\eqref{eq:Coulomb-F-from-Phi} and~\eqref{eq:Coulomb-I-from-Psi}, one obtains the result~\cite{GaspardD2018a}
\begin{equation}\label{eq:Coulomb-H-from-I}
H^\pm_{\eta\ell}(\rho) = \frac{\E^{2\pi(\eta+\I\ell)}-1}{\pi} \left[\frac{1}{w_{\eta\ell}} I_{\eta\ell}(\rho) + h^\pm_{\eta\ell} F_{\eta\ell}(\rho)\right]  \quad\forall\ell\in\tfrac{1}{2}\mathbb{Z}  \:.
\end{equation}
Interestingly, the result~\eqref{eq:Coulomb-H-from-I} explicitly provides the analytic structure of the standard irregular Coulomb functions in the complex $k$ plane.
Indeed, except from the normalization coefficient $C_{\eta\ell}$ of $F_{\eta\ell}$ and $I_{\eta\ell}$, all the singularities of $H^\pm_{\eta\ell}$ in $k$ are gathered in the $\rho$-independent functions $\E^{2\eta\pi}$, $(w_{\eta\ell})^{-1}$, and $h^\pm_{\eta\ell}$.
\par Furthermore, the irregular Coulomb function $G_{\eta\ell}$ exhibits a similar decomposition to Eq.~\eqref{eq:Coulomb-H-from-I}.
The definition~\eqref{eq:Coulomb-G} of $G_{\eta\ell}$ leads to the appearance of the function~\cite{Humblet1984, Humblet1990, Lambert1969, Hamilton1973}
\begin{equation}\label{eq:Bethe-g}
g_{\eta\ell} = \frac{h^+_{\eta\ell} + h^-_{\eta\ell}}{2} = \frac{\psi(\ell+1+\I\eta) + \psi(\ell+1-\I\eta)}{2} - \ln\eta  \:,
\end{equation}
that is real for $\eta>0$.
As shown in Fig.~\ref{fig:map-Bethe-g}(c), the function $g_{\eta\ell}$ displays the Coulomb poles from both $h^+_{\eta\ell}$ and $h^-_{\eta\ell}$ in addition to the logarithmic branch cut.
In Eq.~\eqref{eq:Bethe-g}, the branch cuts of $h^\pm_{\eta\ell}$ --- located on the imaginary axis of $\anbohr k$ due to the principal value of the complex logarithm --- have been turned towards the negative real axis.
The functions $h^\pm_{\eta\ell}$ and $g_{\eta\ell}$ also satisfy the reflection formulas
\begin{equation}\label{eq:Bethe-reflection}
h^\pm_{\eta,-\ell-1} = h^\pm_{\eta,\ell}  \quad\textrm{and}\quad  g_{\eta,-\ell-1} = g_{\eta,\ell}  \quad\forall\ell\in\mathbb{C}  \:.
\end{equation}
Stemming from Eq.~\eqref{eq:Coulomb-w-reflection}, these properties are useful in calculations involving the transformation $\ell\mapsto-\ell-1$.
\par Finally, the analytic decomposition of $G_{\eta\ell}$ reads
\begin{equation}\label{eq:Coulomb-G-from-I}
\boxed{G_{\eta\ell}(\rho) = \frac{\E^{2\pi(\eta+\I\ell)}-1}{\pi} \left[\frac{1}{w_{\eta\ell}} I_{\eta\ell}(\rho) + g_{\eta\ell} F_{\eta\ell}(\rho)\right]   \quad\forall\ell\in\tfrac{1}{2}\mathbb{Z}}  \:.
\end{equation}
This last expression has important consequences in the effective-range theory of charged particle scattering.\cite{GaspardD2018a}
\par The decomposition~\eqref{eq:Coulomb-G-from-I} remains valid for complex valued $\eta$ and $\rho$ except on the branch cut discontinuities of $F_{\eta\ell}$, $I_{\eta\ell}$ and $g_{\eta\ell}$.
Furthermore, Eq.~\eqref{eq:Coulomb-G-from-I} allows us to determine the analytic continuation of the function $G_{\eta\ell}$ in the complex plane of the wave number $k$ and then of $\rho$.
Since the function $g_{\eta\ell}$ shown in Eq.~\eqref{eq:Bethe-g} is proportional to $\ln\rho$, it is defined up to a multiple of $2\pi\I$.
As a consequence of Eq.~\eqref{eq:Coulomb-G-from-I}, the transformation $\rho\mapsto\rho\E^{2\pi\I n}$ for $\ell\in\tfrac{1}{2}\mathbb{Z}$ leads to the novel continuation:
\begin{equation}\label{eq:Coulomb-G-continuation}
G_{\eta\ell}(\rho\E^{2\pi\I n}) = \E^{2\pi\I n\ell}\left[G_{\eta\ell}(\rho) + 2\I n\left(\E^{2\pi(\eta+\I\ell)}-1\right) F_{\eta\ell}(\rho)\right]  \quad\forall n\in\mathbb{Z}  \:.
\end{equation}
Therefore, the corresponding multifunction $\mathbf{G}_{\eta\ell}$ is given by
\begin{equation}\label{eq:Coulomb-G-surface}
\mathbf{G}_{\eta\ell}(\rho) = \left\{\pm G_{\eta\ell}(\rho) \pm 2\I n\left(\E^{2\pi(\eta+\I\ell)}-1\right) F_{\eta\ell}(\rho), ~n\in\mathbb{Z}\right\}  \:.
\end{equation}
Both upper and lower signs in Eq.~\ref{eq:Coulomb-G-surface} have to be included due to the two sheets of the normalization factor $C_{\eta\ell}$ in $I_{\eta\ell}$ and $F_{\eta\ell}$. 
\par The connection formula~\eqref{eq:Coulomb-Psi} between the modified Coulomb functions $\Phi_{\eta\ell}$ and $\Psi_{\eta\ell}$ can be used to relate the standard functions $F_{\eta\ell}$, $H^\pm_{\eta\ell}$ and $G_{\eta\ell}$ between them.  
For this purpose, one has to calculate the derivatives $\dot{\Phi}_{\eta,\ell}$ and $\dot{\Phi}_{\eta,-\ell-1}$ in the right hand side of Eq.~\eqref{eq:Coulomb-Psi}.
We have shown using relation~\eqref{eq:Coulomb-F-from-Phi} that $\dot{\Phi}_{\eta\ell}$ can be written as
\begin{equation}\label{eq:Coulomb-DotPhi-from-F-1}
\dot{\Phi}_{\eta\ell} = \frac{(2\eta)^{\ell+1}}{C_{\eta\ell}\Gamma(2\ell+2)}\left[\dot{F}_{\eta\ell} - g_{\eta\ell}F_{\eta\ell}\right]  \:.
\end{equation}
The function $g_{\eta\ell}$ comes from the logarithmic derivative with respect to $\ell$ of the coefficient in front of the square brackets.
The function $\dot{\Phi}_{\eta,-\ell-1}$ can be obtained from Eq.~\eqref{eq:Coulomb-DotPhi-from-F-1} using the reflection formulas~\eqref{eq:Coulomb-C-reflection-1}, \eqref{eq:Coulomb-F-reflection-2} and~\eqref{eq:Bethe-reflection} 
\begin{equation}\label{eq:Coulomb-DotPhi-from-F-2}
\dot{\Phi}_{\eta,-\ell-1} = \frac{(2\eta)^{\ell+1}w_{\eta\ell}}{C_{\eta\ell}\Gamma(2\ell+2)}\left[\sign(\Re\eta)^{2\ell+1}\dot{F}_{\eta,-\ell-1} - g_{\eta\ell}F_{\eta\ell}\right]  \quad\forall\ell\in\tfrac{1}{2}\mathbb{Z}  \:.
\end{equation}
After some simplifications, the modified Coulomb function $I_{\eta\ell}$ from Eq.~\eqref{eq:Coulomb-I-from-Psi} reads
\begin{equation}\label{eq:Coulomb-I-from-F}
I_{\eta\ell} = \frac{w_{\eta\ell}}{2}\left[\dot{F}_{\eta\ell} + \sign(\Re\eta)^{2\ell+1}\dot{F}_{\eta,-\ell-1} - 2g_{\eta\ell}F_{\eta\ell}\right]  \:.
\end{equation}
The irregular Coulomb functions $H^\pm_{\eta\ell}$ directly follow from the connection formula~\eqref{eq:Coulomb-H-from-I}
\begin{equation}\label{eq:Coulomb-H-from-F}
H^\pm_{\eta\ell} = \frac{\E^{2\pi(\eta+\I\ell)}-1}{2\pi}\left[\dot{F}_{\eta\ell} + \sign(\Re\eta)^{2\ell+1}\dot{F}_{\eta,-\ell-1}\right] \pm\I F_{\eta\ell}  \quad\forall\ell\in\tfrac{1}{2}\mathbb{Z}  \:.
\end{equation}
To get the last term in Eq.~\eqref{eq:Coulomb-H-from-F} out of the square brackets, we have used the property  
\begin{equation}\label{eq:Bethe-h-from-g}
h^\pm_{\eta\ell} = g_{\eta\ell} \pm\frac{\I\pi}{\E^{2\pi(\eta+\I\ell)}-1}  \quad\forall\ell\in\tfrac{1}{2}\mathbb{Z}  \:.
\end{equation}
Finally, the Coulomb function $G_{\eta\ell}$ can be calculated from Eq.~\eqref{eq:Coulomb-H-from-F} by means of the definition~\eqref{eq:Coulomb-G}
\begin{equation}\label{eq:Coulomb-G-from-F}
\boxed{G_{\eta\ell} = \frac{\E^{2\pi(\eta+\I\ell)}-1}{2\pi}\left[\dot{F}_{\eta\ell} + \sign(\Re\eta)^{2\ell+1}\dot{F}_{\eta,-\ell-1}\right]  \quad\forall\ell\in\tfrac{1}{2}\mathbb{Z}}  \:.
\end{equation}
This result is consistent with equation (6.3) of Ref.~\citenum{Dzieciol1999}.
\par It should be noted that the sign appearing in front of $\dot{F}_{\eta,-\ell-1}$ can be formally omitted using the multifunction $\dot{\mathbf{F}}_{\eta\ell}$, as discussed~Sec.~\ref{sec:Coulomb-symmetries}.
However, unlike $\mathbf{F}_{\eta\ell}$, the multifunctions $\dot{\mathbf{F}}_{\eta,\ell}$ and $\dot{\mathbf{F}}_{\eta,-\ell-1}$ display infinitely many Riemann sheets due to the logarithm stemming from the derivative of $\rho^{\ell+1}$ with respect to $\ell$ in the definition~\eqref{eq:Coulomb-F}.
\par The connection formula~\eqref{eq:Coulomb-G-from-F} allows us to determine the action of the transformation $\ell\mapsto-\ell-1$ on the irregular Coulomb function:
\begin{equation}\label{eq:Coulomb-G-reflection}
G_{\eta,-\ell-1}(\rho) = \sign(\Re\eta)^{2\ell+1}\, G_{\eta\ell}(\rho)  \quad\forall\ell\in\tfrac{1}{2}\mathbb{Z}  \:.
\end{equation}
This result can also be obtained from Eq.~\eqref{eq:Coulomb-H-reflection-gen} using the property~\eqref{eq:Coulomb-C-reflection-2}.
The similarity of Eq.~\eqref{eq:Coulomb-G-reflection} with the reflection formula~\eqref{eq:Coulomb-F-reflection-2} for $F_{\eta\ell}$ follows from the two-sheeted structure of $C_{\eta\ell}$ in $\dot{F}_{\eta,\ell}$ and $\dot{F}_{\eta,-\ell-1}$.

\section{Conclusions\label{sec:Conclusions}}
\par In this paper, the connection formulas between the regular and the irregular Coulomb functions have been established in full generality for either complex or  half-integer values of $\ell$. 
For this purpose, we have first discussed the symmetry properties of $F_{\eta\ell}$ based on those of the confluent hypergeometric function $\regm(\alpha,\beta,z)$, especially under the transformation $\ell\mapsto-\ell-1$.
We have shown in Eq.~\eqref{eq:Coulomb-F-reflection-2} that this transformation leaves $F_{\eta\ell}$ unchanged except for a sign which can be interpreted as the consequence of the two Riemann sheets in the normalization coefficient $C_{\eta\ell}$, namely $\pm C_{\eta\ell}$, as explained at the end of subsection~\ref{sec:Coulomb-symmetry-index}.
\par A modified Coulomb function $\Phi_{\eta\ell}$ satisfying the Coulomb equation~\eqref{eq:Coulomb-eq} has been introduced at Eq.~\eqref{eq:Coulomb-Phi}.
Being holomorphic in the wave number $k$, the energy $\energy$, and $\ell$, this function $\Phi_{\eta\ell}$ is more convenient than $F_{\eta\ell}$ to study the analytic properties of Coulomb-related functions in the complex planes of $k$ or $\energy$.
\par When coming to the irregular Coulomb functions $H^\pm_{\eta\ell}$ and $G_{\eta\ell}$, two different cases have been considered:
\begin{enumerate}
\item On the one hand, when $\ell$ is not a half-integer, $H^\pm_{\eta\ell}$ and $G_{\eta\ell}$ are a linear combination of $\Phi_{\eta,\ell}$ and $\Phi_{\eta,-\ell-1}$, as shown in Eq.~\eqref{eq:Coulomb-H-from-Phi-0}, because of the definition~\eqref{eq:Tricomi-U-1} of the Tricomi function $\hypu(\alpha,\beta,z)$.
\item On the other hand, the limit $\ell\rightarrow\tfrac{1}{2}\mathbb{Z}$ can be evaluated from l'Hospital's rule, hence the appearance of the derivatives of $\Phi_{\eta,\ell}$ and $\Phi_{\eta,-\ell-1}$ with respect to $\ell$.
The newly introduced modified Coulomb function $\Psi_{\eta\ell}$ containing these derivatives is defined by Eq.~\eqref{eq:Coulomb-Psi}.
The functions $\{\Phi_{\eta\ell},\Psi_{\eta\ell}\}$ form a couple of linearly independent solutions to Eq.~\eqref{eq:Coulomb-eq} for $\ell\in\tfrac{1}{2}\mathbb{Z}$, that are holomorphic in $k$ and $\energy$.
\end{enumerate}
Concerning the non-holomorphic functions $H^\pm_{\eta\ell}$ and $G_{\eta\ell}$, the connection formulas~\eqref{eq:Coulomb-H-from-I} and~\eqref{eq:Coulomb-G-from-I}---also obtained in Ref.~\citenum{GaspardD2018a}---provide all their singular structures in the complex planes of $k$ and $\energy$.
Most of these singularities are described by the $\rho$-independent functions $h^\pm_{\eta\ell}$ and $g_{\eta\ell}$, which play an important role in scattering theory of charged particles.\cite{Breit1959, Humblet1964, Humblet1990, Hamilton1973, Rakityansky2013, GaspardD2018a}
In addition, the connection formulas have led us to the analytic continuation~\eqref{eq:Coulomb-G-continuation} of $G_{\eta\ell}$ for $\ell\in\tfrac{1}{2}\mathbb{Z}$ as well as the corresponding multifunction~\eqref{eq:Coulomb-G-surface}.
\par Finally, we have directly related $H^\pm_{\eta\ell}$ and $G_{\eta\ell}$ to the derivatives of $F_{\eta\ell}$ with respect to $\ell$ in Eqs.~\eqref{eq:Coulomb-H-from-F} and~\eqref{eq:Coulomb-G-from-F}. 
A consequence of these relations is Eq.~\eqref{eq:Coulomb-G-reflection}, showing that $G_{\eta\ell}$ is unchanged under the transformation $\ell\mapsto-\ell-1$, except for a sign which turns out to be the same as between $F_{\eta,\ell}$ and $F_{\eta,-\ell-1}$.
\par In the future, we plan to use the connection formulas to derive novel representations of $H^\pm_{\eta\ell}$ and $G_{\eta\ell}$ based on those of $\Phi_{\eta\ell}$.
In particular, integral representations and expansions in series of Bessel functions, which are useful to describe the low-energy scattering of charged particles.

\begin{acknowledgments}
The author thanks Professor J.-M. Sparenberg for useful suggestions that contributed to improve the manuscript.
The author acknowledges the financial support of a fellowship granted by the \'Ecole polytechnique de Bruxelles.
This work has also received funding from the European Union's \href{https://doi.org/10.13039/100010662}{Horizon 2020 (Excellent Science)} research and innovation program under Grant Agreement No. 654002.
\end{acknowledgments}

\end{document}